**Data-Driven Prediction Model of Components Shift during Reflow Process in Surface Mount Technology**

Irandokht Parviziomran, Shun Cao, Krishnaswami Srihari, Daehan Won

[a]Department of Systems Science and Industrial Engineering, State University of New York, Binghamton, New York 13902

**Abstract**

In surface mount technology (SMT), mounted components on soldered pads are subject to move during reflow process. This capability is known as self-alignment and is the result of fluid dynamic behaviour of molten solder paste. This capability is critical in SMT because inaccurate self-alignment causes defects such as overhanging, tombstoning, etc. while on the other side, it can enable components to be perfectly self-assembled on or near the desire position. The aim of this study is to develop a machine learning model that predicts the components movement during reflow in $x$ and $y$-directions as well as rotation. Our study is composed of two steps: (1) experimental data are studied to reveal the relationships between self-alignment and various factors including component geometry, pad geometry, etc. (2) advanced machine learning prediction models are applied to predict the distance and the direction of components shift using support vector regression (SVR), neural network (NN), and random forest regression (RFR). As a result, RFR can predict components shift with the average fitness of 99%, 99%, and 96% and with average prediction error of 13.47 (µ$m$), 12.02 (µ$m$), and 1.52 (deg.) for component shift in $x$, $y$, and rotational directions, respectively. This enhancement provides the future capability of the parameters' optimization in the pick and placement machine to control the best placement location and minimize the intrinsic defects caused by the self-alignment.

**1. Introduction**

Surface mount technology (SMT) is an enhanced method in electronic packaging in which surface mount components (SMCs) are placed directly on the printed circuit boards (PCB). Basically, surface mount assembly (SMA) line contains three main operations as stencil printing process (SPP), pick and placement process (P&P), and reflow soldering. At first stage, solder paste is printed onto the pads of PCB to establish electrical connections, followed by picking and placing SMCs on their corresponding pads. Then, reflow oven is employed to attach SMCs by forming permanent solder joints. Basically, solder joint is formed by passing PCB through several thermal zones, known as thermal profile. First, PCB is preheated to start soaking. Then, temperature is increased by defined peak level to melt deposited solder paste. Finally, solder joint is formed by reducing the temperature in cooling zone. The amount of temperature and time that PCB spends at each zone is predefined based on factors, such as type of used solder paste, material and thickness of PCB, number and type of SMCs, etc. Moreover, fluid dynamic behavior of solder paste enforces SMCs to move during reflow process, once solder paste starts melting. Mainly, force originating from the surface tension empowers components to move in a direction that achieves the most magnificent symmetry [1] on the subject of its relative position with soldered pad [2]. This capability is well known as SMCs self-alignment, self-assembled, etc. However, there is no promises that self-alignment always compensates misplacing on P&P process, it can be used to optimize placement machining parameters.

Despite numerous studies investigating the theoretical fundamentals of components movements during reflow [1-8], there is no apparent justification to address practical challenges of the self-alignment in the real situation. On one side, there is no specific rule listed in the literature that recommends which control variables have a significant contribution in components movement in $x$, $y$-directions as well as rotation. On the other side, the widely applied methods are conceptual models along with simulation and numerical models that are accompanied by experimental results [3]. With respect to the privilege of data-driven techniques in compare with the conventional statistical methods [4] , there is a few data-driven model for predicting components movement as well as components position after reflow process. For instance, Lv et al. presented a comprehensive review of the application of data mining techniques in electronic industries [5]. According to their investigation, a few studies address the applied data mining technique in the printing process as well as the reflow process, but none of them studied components self-alignment from big data standpoints [5]. Moreover, a study has been conducted by Marktinek et al. concurrently, which used a neural network model to predict components movement [6]. This study presented a one layer neural network to predict components position after reflow regarding its position before reflow in $x$, $y$, and rotational directions [6] while the importance of solder paste properties such as volume, offset, etc. are neglected. Hence, it enhances the necessities of: (1) obtaining insight into the most significant contributing factors in components shift, (2) outlining machine learning algorithm that predicts components movement during reflow concerning these factors. Moreover, a generalized prediction model that addresses different type and size of passive chip components movement could be a breakthrough of this area since all previous studies considered only one or two types of components [1, 3, 6, 7].

From the extensive research conducted to develop this research [2, 7, 8], and the domain knowledge, 48 variables are selected to train the prediction models (see Fig. 1). These contributing factors are categorized as components geometry, pad geometry, solder paste inspection, placement inspection, paste-pad relative factors, placement-paste relative factors and placement-pad relative factors that are presented in Fig. 1. Furthermore, 3 targets are defined as components shift in $x$, $y$ direction, and rotation that are predicted with the help of remarkable machine learning algorithm, including support vector regression (SVR), deep neural network (NN), and random forest regression (RFR).

The rest of this paper is organized as follows: Section 2 presents the literature related to components self-alignment; the built prediction models are discussed in Section 3 followed by results in Section 4; the conclusions and future work of this research are provided in Section 5.

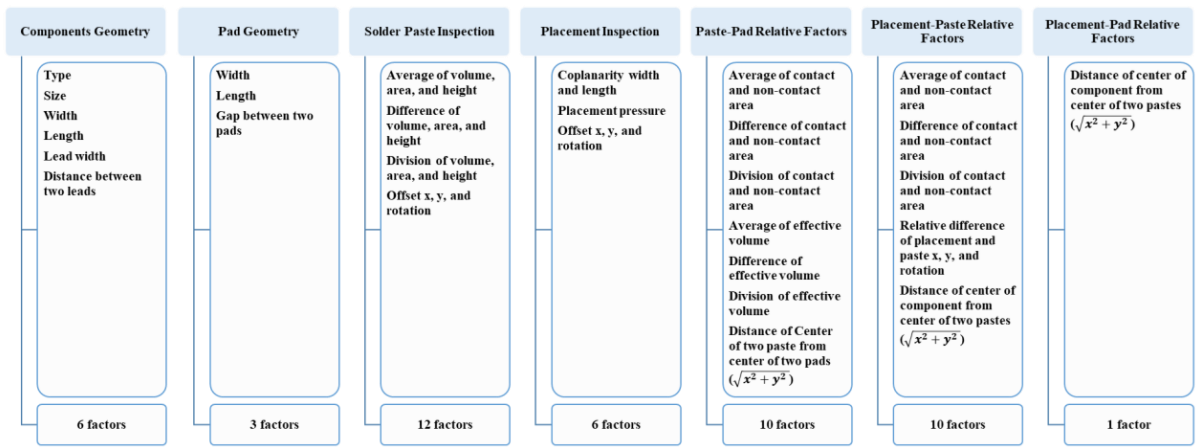

Fig. 1. contributing factors in components movement during reflow process; Notation: average volume, area, height, contact area, non-contact area and effective volume presents average of variable value in respect of pad 1 and pad 2; difference volume, area, height, contact area, non-contact area and effective volume presents difference of variable value on pad 1 subtracted from pad 2; division volume, area, height, contact area, non-contact area and effective volume presents division of variable value on pad 1 divided by pad 2; solder paste and placement offset presents their offsets ($x$, $y$, and rotation) in respect of reference point which is the center of two pads; contact area presents the overlap area of each solder paste with its corresponding pad or overlap area of each solder paste with component; effective volume presents the amount of solder paste volume in respect of contact area.

## 2. Literature review

Many researchers studied components capability in being self-aligned during the reflow process. The main focus of these researches is analyzing experimental results along with simulation and numerical calculation that is built up based on dynamic fluid concepts. However, various studies address the theoretical aspect behind of components shift during reflow, there is not substantial data-driven study in this regard. Ellis and Masada presented a dynamic model in respect of chip capacitor behavior during the reflow process [8]. They studied the effects of various factors (e.g., pad geometry, solder volume, etc.). Presented model is based on calculated forces acting on component and the moment of their actions [8]. Liukkonen et al. studied the theoretical aspect of the self-alignment mechanism and compared components position before and after reflow in both lead-free and tin-lead processes with experimental results [1]. However, they outlined that there was more variation capability of self-alignment in the lead-free process [1], but other contributing factors rather than solder type is neglected from their experiment. Kong et al. predicted the accuracy of soldered flip chip assemblies based on chip mass, die tilt, and solder volume variations and distributions [7]. They applied a regression model that optimizes static equilibrium conditions of a chip and validated their result experimentally. They concluded that the solder volume variation has greater influence on the chip standoff height (i.e., in the $z$-direction) than the chip lateral alignment accuracy (i.e., in $x$ and $y$-direction) [7]. Moreover, Dusek et al. studied components self-alignment in respect of three types of solders (one leaded and two lead-free) that have been deposited with five different angle offsets and two different volumes which is reflowed with two different reflow technologies [2]. Krammer investigated restoring forces that acting on passive chip component (0603) during reflow conceptually which is validated by a simulation model and experimental result [3].

## 3. Methodology

In this section, we discuss machine learning techniques that are applied to predict passive chip components shift after the reflow process. At first, a comprehensive description of the experiment and collected data are presented and contributing factors under this study are introduced. Finally, a brief description of applied machine learning techniques and their justification for this study are presented.

*3.1. Data and experiment description*

The experiment is designed to assemble 6 passive chip component types, including 3 resistors and 3 capacitors in 3 size categories as R1005 and C1005 (1mm×0.5mm), R0603 and C0603 (0.6mm×0.3mm), and R0402 and C0402 (0.4mm×0.2mm). The PCB is designed in which each above-named components are placed on two corresponding pads (named as pad1 and pad 2) that are covered with solder paste in SPP. Fig. 2 (a) shows the description of pads, solder pastes and component in $x$ and $y$ plane. The red dot in the Fig. 2 (a) indicates the center of two pads which is considered as a reference point (0, 0) in this study. Since all components are placed horizontally, longer and shorter sides of the component are considered as parallel with $x$ and $y$ directions respectively, and rotation is defined as a circular movement on $x$-axis (see Fig. 2 (b)). Initially, lead-free solder paste is deposited on the top of pads, concerning paste volume average, paste volume difference, and paste intentional offset ($x$, $y$, rotation) as interest factors in SPP stage. Then, components are placed with mounter machine, considering placement pressure and placement intentional offset ($x$, $y$, rotation) as interest factors in P&P stage. Finally, all components are permanently attached on PCB through nitrogen reflow oven to minimize the effect of PCB's surface oxidation. In detail, 33 combinations of above-mentioned interest factors are considered to place 6 types of passive chip components with 20 replications.

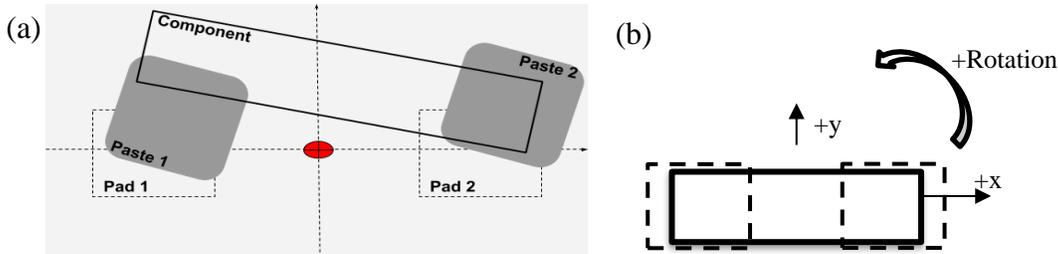

Fig. 2. (a) pads, pastes and component description in 2D (x & y) plane; (b) a definition of the direction of components shift.

During the reflow process, surface tension that is originated from liquid state of solder paste drags component to the center of pads in which component moves in $x$ and $y$-direction and rotates around $x$-axis. Basically, component shift is defined as the relative locational difference between component position before reflow stage and after reflow stage. Note that we consider such amount of the shift as a prediction target for our study. Specifically, we categorized all 48 factors into 7 groups as input features for the predictive modeling analysis (see Fig. 1). Among these 7 categories, the first two is obtained in respect of components directory and PCB design. It is worth mentioning that some detailed component properties like component plating type and thickness, weight, etc. are not considered in our study. Paste inspection and placement inspection are outlined from solder paste inspection (SPI) machine and pre reflow automated optical inspection (Pre-AOI) machine. The last three relative factors are calculated based on aforementioned categories.

## 3.2. Data pre-processing

However, this experiment is designed to place 3940 components in 6 type categories, but some components are missing during the P&P and reflow soldering processes that could be the result of defined placement job file or components floating in reflow stage. Since the cause of missing components is not under the scope of this study, these missing values are eliminated in our study (i.e., we just considered components that are permentley attached after reflow sodering process). Moreover, because of large enough data size, outliers are removed from dataset and gradually, the sample size is reduced to 3917 records with 48 factors. Furthermore, component type is coded as component size for three size categories and resistors and capacitors are distinguished by categorical values -1 and 1 respectively in order to provide a generalized prediction model for all type and size of passive chip components. In respect to dimensionality reduction, The Spearman correlation coefficient is applied to address the nonlinear dependency of continuous and ordinal factors and to remove factors that do not correlate with prediction targets [9].

## 3.3. Support vector regression (SVR)

The general SVR formulation and concepts presented in this study is based on reviewing literatures [10, 11]. For a given training set $\{(x_1, y_1), ..., (x_n, y_n)\} \subset X \times R$ where $X$ denotes the space of input features, the SVR method fits hyperplane $F(X)$ (Eq. (1)) where most data points fall on this plane.

$$F(x) = \langle w, x \rangle + b; \ where \ w \in X, b \in R \tag{1}$$

, where $\langle .,. \rangle$ denotes the dot product of the two vectors. Moreover, upper and lower boundary lines with the distance of $\varepsilon$ and $-\varepsilon$ from defined hyperplane is fitted in which data points that fall within this margin of tolerance range will not be penalized while each point that falls out of this boundary would be penalized with the user defined ratio of $C$ in respect of the amount of non-negative deviations $\xi_i$ and $\xi_i^*$. This constraint is known as ε-insensitive loss function. To solve SVR estimation function, the objective is minimizing the norm of hyperplane weight ($w$) with constrains of boundary lines which is presented in Eq. (2) and (3).

$$\min \quad \frac{1}{2} \| w \|_2^2 + C \sum_{i=1}^{n} (\xi_i + \xi_i^*) \tag{2}$$

$$s.t. \quad \begin{cases} y_i - \langle w, x_i \rangle - b \leq \varepsilon + \xi_i \\ -y_i + \langle w, x_i \rangle + b \leq \varepsilon + \xi_i^* \\ \xi_i, \xi_i^* \geq 0 \end{cases} \tag{3}$$

The notable breakthrough of the SVR is using support vectors that makes the model independent from the dimensionality of input space $X$. It enables the algorithm to solve nonlinear problem efficiently. In this study, ε-insensitive loss function [11] is considered along with linear kernel to address any potential linear relation among feature space and target values. Basically, we set as $\varepsilon = 0.1$ defining the distance of boundaries from hyperplane as well as $C = 1$ for the penalty ratio.

## 3.4. Deep neural network (NN)

For a given training set $\{(x_1, y_1), ..., (x_n, y_n)\} \subset X \times R$, independent variable $X$ denotes the space of input and dependent variable $y$ denotes output of a network that are connected through a graph shape path with one or more hidden layers and corresponding hidden neurons. Each input neuron that is representative of a feature, carries an initialized weight in which summation of all input neurons and their corresponding weights will be activated on a hidden neuron with the mean of a predefined activation function [12]. Furthermore, the NN will be fed sequentially and its performance will be evaluated in respect of a predefined loss function. Input weight and bias would be updated by first derivation of activation function (i.e., backpropagation) at each sequence till desire termination condition would be achieved. Estimation function of NN is mentioned in $F(X, W) = \phi(W^T X) + b$, where $b$ is a constant and $\phi$ is corresponding activation function presented as $\phi(W^T X) = \phi(\sum_{i=1}^{n} (w_i x_i))$. Normal initialized weight is considered for the proposed NN architecture, which is activated by rectifier function; $\phi(X) = max(0, x_i)$. Finally, adaptive moment (Adam) optimizer is applied to optimize the associated weigh and bias of training set in respect of mean square error (MSE) of prediction and actual value of $y$. Two hidden layers are proposed for this study along with feature size and 100 hidden neurons respectively.

*3.5. Random forest regression (RFR)*

Random forest is built based on hierarchical tree-structured predictors in which each tree within forest is trained on a random split of input features [13]. Through the top to dawn path from tree's root node to splitting nodes and finally to leaves nodes, initially, algorithm uses all input features. These features would be divided at each stage till finally, the best subset of features would be remained on terminal leaves. In this regard, (1) prediction and feature selection are considered at same time. (2) dimensionality of feature space will not affect prediction model. The general RFR formulation and concepts used in this study is based on reviewing literature [13]. For a given training set $\{(x_1, y_1), ..., (x_n, y_n)\} \subset X \times R$, where $X$ denotes the space of input features, the estimation function of RFR is: $F(x) = \frac{1}{J}\sum_{j=1}^{J} f_j(x)$, where $j$ is the number of trees in forest and $f_j(x)$ is estimation function of each tree. For a decision tree with $M$ splitting nodes, the feature space would be splitting into $M$ regions as $R_m$ in which $f_j(x) = \sum_{m=1}^{M} b_m \varphi(x, R_m); for\ j = 1, ..., J$. A binary decision $\varphi(x, R_m) \in \{0,1\}$ is made to outline whether input feature $x$ is selected (i.e., $x \in R_m$) or not. In this study, the number of trees is initialized as 1,000 and trees are fully grown which means only one leaf is remained at terminal with selected feature space.

## 4. Experimental results

The coefficient of determination ($R^2$) is employed to measure the fitness of learning algorithm on a training set along with root-mean-squared error (RMSE) to evaluate the performance of prediction models on the testing set. As it is mentioned in section 3.1, three continuous variables are defined as targets, named as Shift $X$, Shift $Y$, and Shift Rotation, in which prediction models are trained on each target separately. RFR uses entire training samples in respect of mean-squared error (MSE) as splitting quality measurement. The best split in RFR is determined by searching all potential splits on each input variable, and the decision tree is fully grown. The accuracy of the prediction models and fitness of learning algorithm are evaluated through 10-fold cross validation to (1) ensure the model is well generalized regarding any independent data set; (2) limit the problem of overfitting. Table 1 presents the average and standard deviation of 10-fold RMSE and $R^2$ for the SVR, NN, and RFR prediction models, respectively.

Table 1. performance measurement for SVR, NN, and RFR prediction models; the bolded number represents the outperformed model in terms of both average and standard deviation of 10-fold cross validation.

| Prediction Model | Shift $X$ ($\mu m$) | | | | Shift $Y$ ($\mu m$) | | | | Shift Rotation (deg.) | | | |
|---|---|---|---|---|---|---|---|---|---|---|---|---|
| | RMSE | | $R^2$ | | RMSE | | $R^2$ | | RMSE | | $R^2$ | |
| | Avg. | Std. | Ang. | Std. | Avg. | Std. | Ang. | Std. | Avg. | Std. | Ang. | Std. |
| SVR | 15.29 | 0.86 | 0.95 | 0.00 | 15.23 | 0.79 | 0.89 | 0.00 | 1.58 | 0.07 | 0.68 | 0.00 |
| NN | 12.48 | 1.00 | 0.98 | 0.00 | 11.15 | 0.79 | 0.96 | 0.01 | 1.34 | 0.28 | 0.90 | 0.04 |
| RFR | **13.47** | **0.58** | **0.99** | **0.00** | **12.02** | **0.40** | **0.99** | **0.00** | **1.52** | **0.07** | **0.96** | **0.00** |

To compare the performance of models, average and standard deviation of $R^2$ and RMSE are considered concurrently. In terms of training fitness ($R^2$), all three models are fitted well for Shift $X$ and $Y$ (see average and standard deviation $R^2$ for Shift $X$ and $Y$ in Table 1) with above 90% average and zero standard deviation $R^2$. In respect of shift rotation, NN is fitted with 90% average and 4% standard deviation $R^2$ while RFR outperforms with 96% average and zero standard deviation of fitness. As a conclusion, RFR is better fitted on training dataset for all three targets. On the other side, to consider overfitting of prediction models, average and standard deviation of RMSE are outlined for testing dataset. However, NN performed better in respect of average RMSE for all three target variables but its highest dispersion around average RMSE makes it unstable comparing with other models. Moreover, RFR provides better performance concerning average and standard deviation RMSE of testing dataset in which predicts passive chip components shift during reflow soldering whit 13.47 ($\mu m$), 12.02 ($\mu m$), and 1.52 (deg.) error in $x$, $y$, and rotational direction for this dataset.

Table 2. prediction performance by components type

|  | Shift X (µm) | | | Shift Y (µm) | | | Shift Rotation (deg.) | | |
|---|---|---|---|---|---|---|---|---|---|
|  | SVR | NN | RFR | SVR | NN | RFR | SVR | NN | RFR |
| C1005 | 19.77 | 16.77 | 14.4 | 20.09 | 18.4 | 17.6 | 0.89 | 0.95 | 0.83 |
| R1005 | 22.13 | 20.9 | 19.63 | 15.24 | 11.83 | 11.13 | 1.33 | 1.11 | 1.09 |
| C0603 | 14.56 | 11.34 | 12.31 | 12.94 | 9.28 | 9.8 | 1.33 | 1.39 | 1.15 |
| R0603 | 12.19 | 10.14 | 11.87 | 10.27 | 7.23 | 7.98 | 1.58 | 1.4 | 1.45 |
| C0402 | 10.56 | 6.61 | 7.16 | 13.62 | 8.04 | 12.25 | 2.1 | 1.62 | 2.0 |
| R0402 | 10.25 | 10.1 | 9.68 | 10.24 | 8.47 | 9.48 | 1.79 | 1.35 | 1.78 |

In addition, RMSE for 6 proposed passive chip components are extracted from the testing set, and the performance of the prediction model (RMSE) is outlined for each of them separately presented in Table 2. It is noteworthy that components motion in $x$ and $y$-directions is sensitive to the size of the component. Along with the increment of component size, the prediction error also increases except for C0402 and R0402 in $y$-direction. Moreover, the prediction error is smaller for capacitors in compare with their equivalent resistors except for C0603 and R0603 in $x$-direction while in $y$-direction, prediction error is higher for the case of capacitors. In terms of rotation, compoents rotate more along with the deduction of component size. Additionally, the shift rotation prediction error is smaller for capacitors in compare with their equivalent resistors except for C0402 and R0402. This can be caused by the effect of components body mass on the components' rotation.

In the matter of feature selection, the RFR algorithm derived 4, 6, and 35 factors that contributes in shift $X$, $Y$, and rotation respectively. All placement offsets ($x$, $y$, and rotation) appeared to have the most contribution in RFR prediction followed by average volume for shift $X$, paste-pad average non-contact area for shift $Y$, and pre-paste rotation, pre-paste average contact area and so forth for rotational shift. In Fig. 3, prediction model is compared with actual movement value in $x$, $y$, and rotational directions. As it is shown in Fig. 3 (a), the volume of deposited solder paste affects components shift in $x$ direction. Moreover, when solder pastes are well aligned with their corresponding pads (i.e. small paste-pad non-contact area), components would not move dramatically in $y$ direction (see Fig. 3 (b)). In respect of rotation, components rotate almost equal with their initial placement rotational offset in reverse direction (see Fig. 3 (c)) to achieve near zero rotational offset after reflow process.

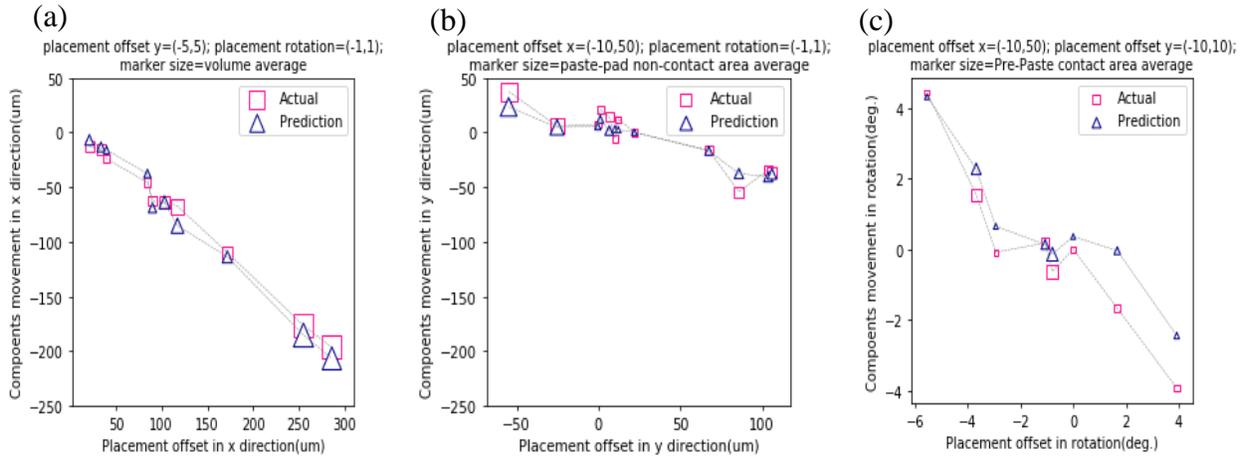

Fig. 3. prediction vs. actual value of components move in (a) x direction; (b) y direction; (c) rotation; blue triangle presents prediction and pink rectangle presents actual value in data set.

## 5. Conclusion and Future work

This study outlined 48 potential factors that contribute in components movement during reflow soldering and applied machine learning approaches such as SVR, NN, and RFR to predict the amount and direction of components move in $x$, $y$ and rotational directions. Based on the result, it is outlined that RFR outperforms other techniques in terms of the average predictive losses and their standard deviations across 10-fold cross validation. Indeed, it is noteworthy that the RFR is well fitted for all 3 targets with the average $R^2$ of 99%, 99%, and 96% with average prediction errors of 13.47 (μm), 12.02 (μm), and 1.52 (deg.) for each shift $X$, $Y$, and rotation respectively. Furthermore, solder paste average volume, paste-pad average non-contact area, and placement-paste average contact area are shown to be important in components shift in $x$, $y$ and rotation respectively. This highlights the importance of solder paste status, relative offset of paste from pad, and relative offset of placement from paste in explaining self-alignment capability of passive chip components.

As a future work, since this study was developed based on the exact value of the component position, defining a standard positional tolerance in respect of component and pad size can enhance the model. Besides, introduced factors at this study can be applied to predict specific defect during reflow soldering such as overhanging, tombstoning, etc. Additionally, the parameters optimization model can be developed to control the best placement location regarding any given solder paste status by the means of the proposed prediction model.